\documentclass[sigconf, natbib=false, nonacm]{acmart}
\AtBeginDocument{%
  }

\usepackage{url}

\usepackage[english]{datetime2}
\usepackage{float}
\usepackage{array}
\usepackage{multicol}

\setcopyright{none}

\RequirePackage[
  datamodel=acmdatamodel,
  ]{biblatex}

\begin{document}

\title{RANDAO-based RNG: Last Revealer Attacks in Ethereum 2.0 Randomness and a Potential Solution}

\author{Do Hai Son}
\email{dohaison1998@vnu.edu.vn}
\affiliation{%
  \institution{VNU Information Technology Institute}
  \country{Hanoi, Vietnam}
}


\author{Tran Thi Thuy Quynh}
\email{quynhttt@vnu.edu.vn}
\affiliation{%
  \institution{University of Engineering and Technology}
    \country{Hanoi, Vietnam}
}

\author{Le Quang Minh}
\email{quangminh@vnu.edu.vn}
\affiliation{%
  \institution{VNU Information Technology Institute}
  \country{Hanoi, Vietnam}
}

\renewcommand{\shortauthors}{Do Hai Son and et al.}

\begin{abstract}
  Ethereum 2.0 is a major upgrade to improve its scalability, throughput, and security. In this version, RANDAO is the scheme to randomly select the users who propose, confirm blocks, and get rewards. However, a vulnerability, referred to as the `Last Revealer Attack' (LRA), compromises the randomness of this scheme by introducing bias to the Random Number Generator (RNG) process. This vulnerability is first clarified again in this study. After that, we propose a Shamir's Secret Sharing (SSS)-based RANDAO scheme to mitigate the LRA. Through our analysis, the proposed method can prevent the LRA under favorable network conditions.  
\end{abstract}




\maketitle

\section{Introduction}

Launching in 2016, Ethereum rapidly rose to rank 2nd in terms of crypto market capitalization~\cite{marketcap} due to its superior features, e.g., smart contract, lower block confirmation time, and so forth. Hence, this network is not only a digital asset but also a decentralized platform for other impactful applications~\cite{Son2021, Lin2022}. However, an intensive number of applications brings the need to scale the Ethereum network and reduce the energy used for mining~\cite{IEEESpectrum2019} through the old consensus mechanism, i.e., Proof-of-Work (PoW). Thus, in 2018, the Ethereum Foundation introduced the Ethereum 2.0 road map~\cite{Ethereumb} to upgrade from PoW to Proof-of-Stake (PoS) using Casper consensus~\cite{Buterin2020}. In 2020, the first phase on the road map, named `Beacon chain', was officially merged into the Ethereum mainnet. After three years, the first phase is done with the highlight, which is a fully PoS-based Ethereum network.

The move from PoW to PoS means the mining process is replaced by stakers, who staked their money to become a validator. Any validator has the chance to propose a new block, a.k.a proposer, and get a bounty of ETH (Ethereum native token). In this work, we focus on the weakness of the random process to select the proposer in Ethereum. According to Eth2.0 specs, this proposer is randomly chosen by the RANDAO scheme~\cite{Ethereumb} (Random decentralised autonomous organisation). Generally, RANDAO is based on CRS (Commit reveal scheme)~\cite{lastreveal}. In CRS, each member commits their value to a group, and the final number is a combination of these values. However, in~\cite{Paul2018}, the author indicated a vulnerability of the RANDAO scheme named LRA. This weakness allows attackers to bias the RNG output. In~\cite{lastreveal}, Vitalik (Ethereum Founder) pointed out that if an attacker had 36\% of the total staked money, he would gain control of the Ethereum network. This means the attacker manipulates the proposers always to be his validators. That leads to the consequence, i.e., he has the ability to validate any invalid blocks. The LRA is similar to the 51\% attack in Ethereum 1.0, but a massive amount of money replaces the overwhelming computational power.

To deal with this vulnerability, the author in~\cite{Boneh2018} proposed the VDF (Verifiable Delay Function) algorithm. This method prevents any validator from finding the final random number before the reveal phase. 
Following that, other studies~\cite{Pietrzak2019, Wesolowski2019, Ephraim2020} proposed their versions of VDF, i.e., simple VDF, efficient VDF, and continuous VDF, respectively. 
The Ethereum Foundation confirmed the VDF version, named minimal VDF~\cite{mVDF}, which will be used in the Ethereum mainnet after phase 2 on the road map. However, the drawback of VDF is that this algorithm requires a specific hardware named `Rig'. Although trustful organizations centrally control this device, it has lost the decentralized properties of blockchain technology. In the worst case, none of `Rig' is available, the VDF is turned down, and then Ethereum will switch back to using the RANDAO scheme with the LRA weakness. In this work, we propose to use the SSS~\cite{Shamir1979} algorithm for the RANDAO process on Ethereum or other Ethereum variants. SSS is an old but effective algorithm for securely sharing a secret. Thus, SSS-based RANDAO can provide a controllable security level to select proposers randomly. In Table~\ref{tab:notation}, we explain several Ethereum terms used in the following sections.

\begin{table}[!b]
\centering
\caption{Ethereum terms used in the paper}
\label{tab:notation}
\resizebox{\linewidth}{!}{
\begin{tabular}{|l|l|} 
\hline
\multicolumn{1}{|>{\centering\hspace{0pt}}m{0.25\linewidth}|}{\textbf{Term}} & \multicolumn{1}{>{\centering\arraybackslash\hspace{0pt}}m{0.685\linewidth}|}{\textbf{Description}} \\ 
\hline
Validator & Instead of `miner' in PoW \\ 
\hline
Proposer & The chosen validator to propose blocks \\ 
\hline
Committee & A set of validators to validate the proposed blocks \\ 
\hline
Effective Balance & The current balance of a validator \\ 
\hline
Slot & Fixed at 12 seconds and equal to a block \\ 
\hline
Epoch & Consists of 32 slots \\
\hline
\end{tabular}
}
\end{table}

Our main contribution in this paper is to propose an SSS-based RANDAO scheme for Ethereum 2.0. In this study, our proposal is a preliminary idea with some security analysis to deal with the LRA attack on Ethereum.

The remainder of the paper is organized as follows. In Section~\ref{sec:LRA}, the brief about LRA attack in RANDAO-based RNG is presented. In Section~\ref{sec:SSS}, we propose to use Shamir's Secret Sharing algorithm for randomly chosen proposers on Ethereum. Finally, Section~\ref{sec:Conclusion} concludes the paper.

\section{Last revealer attacks in RANDAO-based RNG}
\label{sec:LRA}

\begin{figure}
    \centering
    \includegraphics[width=\linewidth]{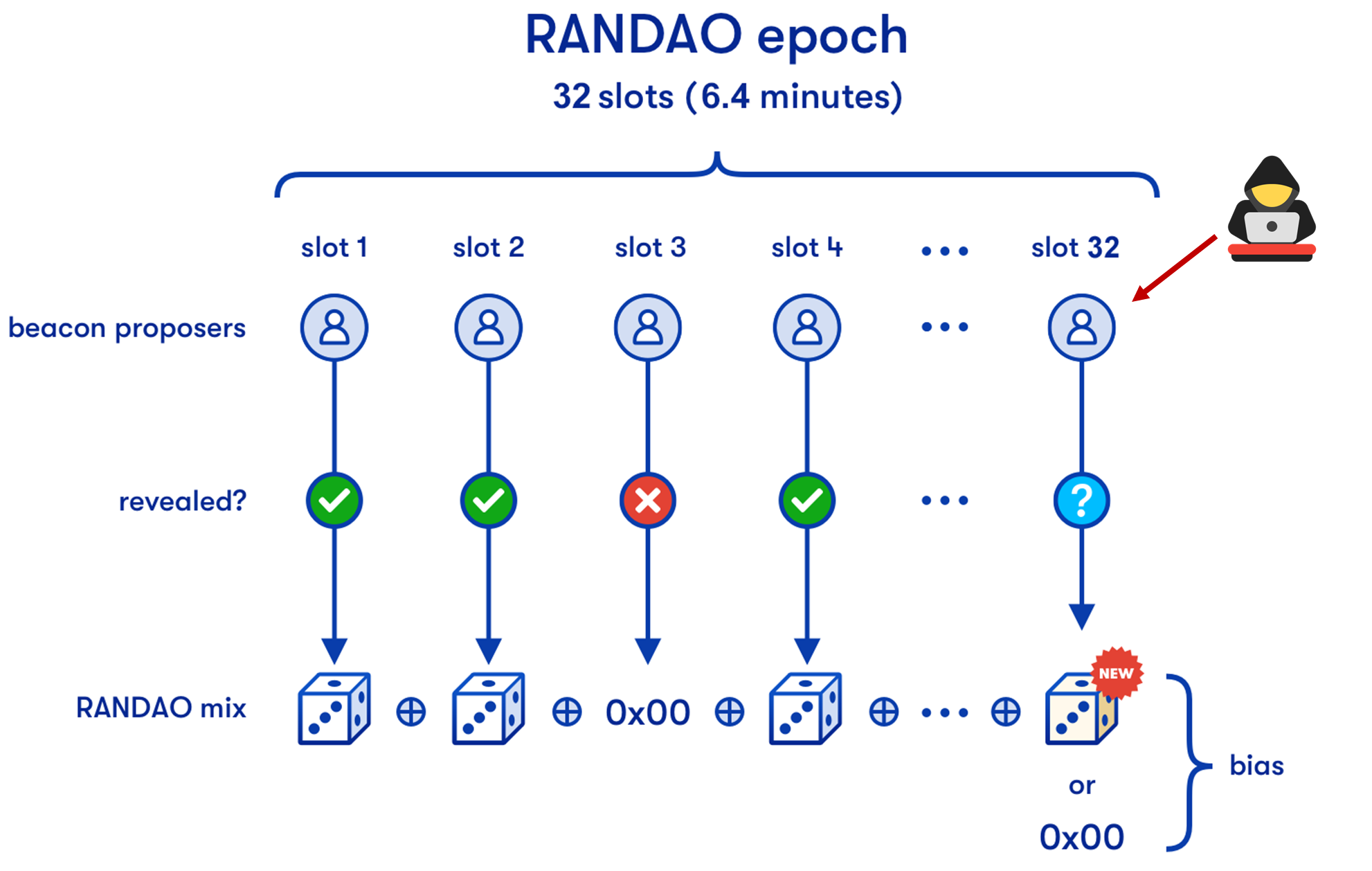}
    \caption{Last revealer attacks on Ethereum.}
    \label{fig:LRA}
\end{figure}

First, we briefly present the RANDAO-based RNG algorithm of Ethereum with an illustration in Fig.~\ref{fig:LRA}. On Ethereum, the set of proposers and committees in ($j+2$)-th epoch is selected from the RANDAO-based results at ($j$)-th epoch. At the first slot in $j$-th epoch, the proposer signs its BLS private\_key on the array of current epoch number ($j$) and fixed DOMAIN\_RANDAO (`0x02000000'). The output of this process is a digital signature ($r_1$) in size of $256$~bits that any validator can verify by the public\_key of the first proposer. This sequence of bits is included in the first slot in $j$-th Epoch. Similarly, the second proposer computes its $256$-bits ($r_2$). However, it does not immediately add this sequence to the block but computes
\begin{equation}
    r_2 = r_1 \oplus r_2,
\end{equation}
where $\oplus$ is the `XOR' operator. The second proposer puts the computed $r_2$ into its block. In the last slot, the sequence is as follows:
\begin{equation}
    r_{32} = \oplus_i r_i, \quad 1 \le i \le 31.
\end{equation}
In the absence of a proposer case, the sequence is preserved as xor with a series of zeros. The final sequence $r_{32}$ together with DOMAIN\_BEACON\_PROPOSER (`0x00000000') and current epoch number ($j$) makes an array called `seed32'. This array is passed through a hash function and returns an array of 32 elements. The final array is one of the two inputs to randomly select the set of proposers for ($j+2$)-th epoch. The latter input is a set of all activated validators in the network at the first slot of $j$-th epoch. The inputs are put into a deterministic function to select the set of validators. Note that there is no randomness in this final validator selection function. Hence, the randomness only comes from the 'seed32' array, which is contributed by a set of proposers.

In the LRA, suppose the attacker is the proposer holder in the 32-th slot. At this stage, he already has information on the set of activated proposers from the first slot and the mixed sequence ($r_{31}$) from the 31 slots previously. Thus, he has two options, i.e., commit his digital signature as usual or pretend to be absent to keep the $r_{31}$. This is a bias in the RNG process. Note that only the last proposer can have this bias because the before validators are infeasible to predict the after's digital signature. That leads to the name `Last revealer attack'. Furthermore, if the attacker has more than one proposer in slots at the tail of an epoch, e.g., $h$ proposers, he will have a total of $2^h$ options to decide whether to propose or not propose a block. Consequently, these $2^h$ choices may introduce biases in the selection of proposers and committees for the ($j+2$)-th epoch.

\section{Shamir's Secret Sharing-based RANDAO}
\label{sec:SSS}
In this section, we present our proposal for SSS-based RNG to prevent the LRA attack. After that, the security analysis and limitations of the proposed scheme are provided.

\subsection{Proposal}

Our proposal is to apply the SSS algorithm to replace the CRS in RANDAO, as mentioned in the previous section. SSS~\cite{Shamir1979} threshold scheme based on polynomial interpolation over finite fields. There are three definitions of SSS ($n, m$), as follows:
\begin{itemize}
    \item Secret ($\mathbf{s}_k$): is the secret information (e.g., number, text, etc.) that needs to be shared securely.
    \item Share: is a part of the secret. For example, from an $\mathbf{s}_k$, we have $m$ shares, i.e., $\mathbf{s}_k 1$, $\mathbf{s}_k 2$,~\ldots, $\mathbf{s}_k n$,~\ldots, $\mathbf{s}_k m$. 
    \item Threshold ($n$): is the minimum number of shares required to recover the original secret ($\mathbf{s}_k$).
\end{itemize}

\begin{figure*}[!ht]
    \centering
    \includegraphics[width=.7\linewidth]{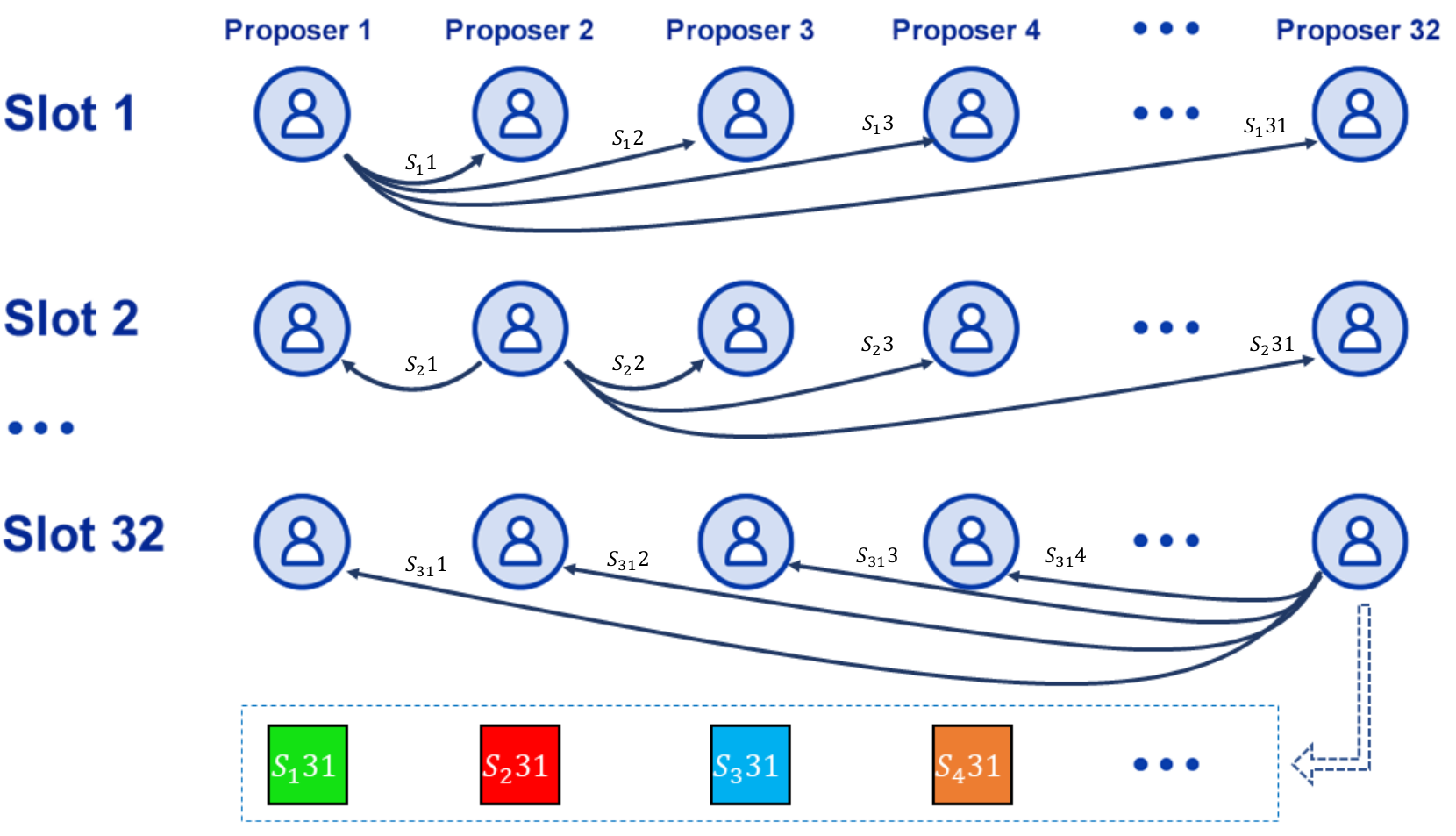}
    \caption{SSS-based RANDAO.}
    \label{fig:SSS}
\end{figure*}

As shown in Fig.~\ref{fig:SSS}, at the first slot, the proposer still computes its digital signature ($r_1$), then divides $r_1$ into 31 shares ($m=31$) corresponding to the rest of proposers in this epoch, i.e., $\mathbf{s}_1 1, \mathbf{s}_1 2,~\ldots, \mathbf{s}_1 31$, respectively. These shares are encrypted using asymmetric key schemes, e.g., RSA~\cite{RSA}, so only the validator identified by its public key can reveal the corresponding share. The first proposer then packs these shares into its block and broadcasts the block to the network. The rest of the proposers also do the same in their slots. At the last proposer, what it has is 31 shares (maximum), i.e., $\mathbf{s}_1 31, \mathbf{s}_2 31,~\ldots, \mathbf{s}_{31} 31$. Using these shares, the last validator cannot recover any digital signature ($r_i$) from the rest of the proposers. This implies that it is not feasible for this validator to manipulate the RNG process of RANDAO. After 32 slots (an epoch), there is the reveal phase, when validators broadcast their decrypted shares. Each secret from $i$-th proposer ($\mathbf{s}_i$) requires at least $n$ shares ($n$ proposers broadcast their $\mathbf{s}_k i, \; 1 \le i \le n$, decrypted shares) to recover. The unrecoverable $r_i$, due to not having enough shares, can be treated as absent proposers, as RANDAO did. RANDAO scheme selects proposers and committees for the $(j+2)$-th epoch from recovered digital signatures and the set of activated validators. The reveal and mixing phase in SSS-based RANDAO occurs after an epoch has been completed and may lead to a delay in the timeline of Ethereum. Inspired by minimal VDF~\cite{mVDF}, the reveal and mixing phase can be conducted in parallel as a pipeline process.

\subsection{Security analysis}

Assume that $t, h$ is the number of proposers who joined SSS-based RANDAO and the number of dishonest proposers in an epoch, respectively. In the first case, as follows:
\begin{equation}
    t \ge n \quad \textbf{and} \quad h < n,
\end{equation}
The proposed method can recover all secrets, and the attacker cannot predict the digital signatures of other honest proposers. Thus, the LRA is prevented. In the second case, as follows:
\begin{equation}
    t < n,
\end{equation}
This is a bad case when we do not have enough shares to recover any secret. This leads to no `seed32' array, the original RANDAO mechanism will break, and a backup solution is needed for this case. In the third case, when attackers have a few proposers in an epoch, as follows:
\begin{equation}
    t \ge n \quad \textbf{and} \quad h \ge n,
\end{equation}
Attackers have enough shares to recover secrets and compute the `seed32' array before starting the reveal phase. This leads to an attack similar to LRA, but the attackers' proposers do not need to be at the tail of the epoch.

The controllable property of the proposed scheme is that we can adapt the threshold $n$ to guarantee the randomness of RANDAO-based RNG. However, it is a trade-off because a higher $n$ means more proposers must be present in the reveal phase.

\section{Conclusion}
\label{sec:Conclusion}

In this work, we proposed to use an SSS-based RANDAO scheme instead of CRS to mitigate the LRA. We first represented the RANDAO-based RNG of Ethereum and pointed out its weakness. The SSS algorithm is applied to RANDAO and shows its advantages and limitations. Our study is still in its preliminary stages and requires further analysis to estimate the amount of stake attackers need to control the mechanism we have proposed.


\printbibliography

\end{document}